\documentclass[fleqn,usenatbib]{mnras}

\usepackage{newtxtext,newtxmath}
\usepackage{xcolor}
\usepackage[T1]{fontenc}

\DeclareRobustCommand{\VAN}[3]{#2}
\let\VANthebibliography\thebibliography
\def\thebibliography{\DeclareRobustCommand{\VAN}[3]{##3}\VANthebibliography}

\usepackage{graphicx}	
\usepackage{multirow}
\usepackage{bm}
\usepackage{hyperref}
\usepackage{nameref}

\newcommand{\xb}{\ensuremath{\boldsymbol{x}}}
\newcommand{\yb}{\ensuremath{\boldsymbol{y}}}

\newcommand{\nb}{\ensuremath{\boldsymbol{n}}}

\newcommand{\eC}{\mathbb{C}}
\newcommand{\eR}{\mathbb{R}}
\newcommand{\eN}{\mathbb{N}}

\title[AIRI-ASKAP]{Scalable precision wide-field imaging in radio interferometry: \\ II. AIRI validated on ASKAP data}

\author[A. G. Wilber et al.]{
A. G. Wilber,$^{1}$
A. Dabbech,$^{1}$
M. Terris,$^{1}$
A. Jackson,$^{2}$
Y. Wiaux$^{1}$\thanks{E-mail: y.wiaux@hw.ac.uk}
\\
$^{1}$Institute of Sensors, Signals and Systems, Heriot-Watt University, Edinburgh EH14 4AS, UK\\
$^{2}$EPCC, The University of Edinburgh, Edinburgh EH8 9BT, UK\\
}

\date{Accepted XXX. Received YYY; in original form ZZZ}

\pubyear{2023}

\begin{document}
\label{firstpage}
\pagerange{\pageref{firstpage}--\pageref{lastpage}}
\maketitle

\begin{abstract}
 
Accompanying Part I, this sequel delineates a validation of the recently proposed AI for Regularisation in radio-interferometric Imaging (AIRI) algorithm on observations from the Australian Square Kilometre Array Pathfinder (ASKAP). The monochromatic AIRI-ASKAP images showcased in this work are formed using the same parallelised and automated imaging framework described in Part I: ``uSARA validated on ASKAP data''. Using a Plug-and-Play approach, AIRI differs from uSARA by substituting a trained denoising deep neural network (DNN) for the proximal operator in the regularisation step of the forward-backward algorithm during deconvolution. We build a trained shelf of DNN denoisers which target the estimated image-dynamic-ranges of our selected data. Furthermore, we quantify variations of AIRI reconstructions when selecting the nearest DNN on the shelf versus using a universal DNN with the highest dynamic range, opening the door to a more complete framework that not only delivers image estimation but also quantifies epistemic model uncertainty. We continue our comparative analysis of source structure, diffuse flux measurements, and spectral index maps of selected target sources as imaged by AIRI and the algorithms in Part I -- uSARA and {\tt WSClean}. Overall we see an improvement over uSARA and {\tt WSClean} in the reconstruction of diffuse components in AIRI images. The scientific potential delivered by AIRI is evident in further imaging precision, more accurate spectral index maps, and a significant acceleration in deconvolution time, whereby AIRI is four times faster than its sub-iterative sparsity-based counterpart uSARA.
\end{abstract}




\section{Introduction}
\label{sec:intro}

The superior detection capabilities of modern and upcoming radio arrays -- namely, the Square Kilometre Array (SKA) -- necessitate new and improved radio-interferometric (RI) imaging algorithms that are precise, robust, and scalable to larger quantities of data, wider fields-of-view, and broader frequency bandwidths. The ``Scalable precision wide-field imaging in radio interferometry'' series aims to showcase a novel imaging framework in action, as applied to real radio observations from ASKAP. The proposed imaging framework builds from compressed sensing techniques to reconstruct true signal from incomplete, noisy data and operates at the interface of optimisation theory and deep learning. Implemented in MATLAB, the framework is automated, highly parallelised, and capable of producing wide-field, high-dynamic range, super-resolved monochromatic intensity images. Two interchangeable image regularisation denoisers can be ``plugged'' in as the `backward' step of the underlying iterative forward-backward (FB) deconvolution structure \citep{Terris22,dabbech2022first} of the imaging framework, alternating with a gradient descent `forward' step promoting data fidelity. 

The first algorithm, unconstrained Sparsity Averaging Reweighted Analysis (uSARA), is purely optimisation-based. It leverages the proximal operator of a state-of-the-art handcrafted sparsity-promoting regularisation function  \citep{Carrillo12,Terris22} as regularisation denoiser. In Part I: ``uSARA validated on ASKAP data'', uSARA was validated against the widely-used CLEAN-based imager for RI {\tt WSClean} \citep{2014MNRAS.444..606O, 2017MNRAS.471..301O}. Our experiments with uSARA showed that we were able to take wide-field, imperfectly calibrated data and create images with exceptional resolution and enhanced sensitivity. The findings of Part I establish uSARA as an advanced RI imaging algorithm capable of surpassing the state-of-the-art in precision and robustness when applied to large-scale, real, and imperfect data. A remaining caveat with uSARA lies in its computational cost due to the iterative nature of the proximal operator underlying its image model. 

In this sequel, we showcase AIRI \citep{Terris22} -- the second algorithm encapsulated in our automated, parallelised imaging framework -- which combines optimisation theory with the power of AI. Building on the recent success of Plug-and-Play (PnP) approaches in various applications such as image restoration \citep{zhang2020plug} and magnetic resonance imaging \citep{ahmad2020plug}, AIRI relies on the same FB iterative scheme as uSARA, with the proximal operator enforcing the image prior model replaced by a learned DNN denoiser. The learned denoiser, deployed on graphic processing units (GPUs), enables a significant acceleration of the backward step of the iterative FB algorithm when compared to the computationally heavy iterative proximal operator powering uSARA. The speed combined with the learning power of the DNNs -- trained herein as denoisers for high dynamic-range images -- make for a scalable and robust tool in image reconstruction. Although the training of the DNNs requires important computational time and resources, AIRI denoisers are pre-trained independently of the data under scrutiny. They can generalise to any RI data with a simple scaling procedure.

This paper specifically addresses testing AIRI on the same real and imperfectly calibrated data from ASKAP, used in Part I. To summarise from Part I, we selected three fields-of-view (FoVs) from ASKAP Early Science and Pilot Survey observations hosting radio sources of primary interest that exhibit emission with both complex diffuse and compact filamentary morphology. Our targets of interest include the merging galaxy cluster system Abell 3391-95 (hosting a candidate radio phoenix; \citealp{2021A&A...647A...3B}), the merging galaxy cluster SPT-CL J2023-5535 (hosting a radio halo and radio relic; \citealp{2020ApJ...900..127H}), the X-shaped radio galaxy PKS 2014-558 \citep[e.g.][]{2020MNRAS.495.1271C}, and the ``dancing ghosts,'' known collectively as PKS 2130–538 \citep[e.g.][]{2021PASA...38...46N}. We refer the reader to Table 1 of Part I for full details on the ASKAP observations selected for imaging. Further details of the observations selected, calibration and processing of the data, and the steps involved to prepare for imaging are elaborated upon in Section 3 of Part I. 

The remainder of this article is structured as follows. In Section~\ref{sec:methods}, we recall the parallelised, automated imaging framework from Part I and expand upon the application of the AIRI algorithm as well as our approach to selecting DNN denoisers for imaging. In Section~\ref{sec:data}, we provide details of the ASKAP data used in this work and the imaging settings applied for the AIRI algorithm. Reconstruction results of our selected fields are presented and compared to the results of Part I in Section~\ref{sec:results}.  The computational performance of AIRI is studied in \ref{sec:time}. Finally, conclusions are made in Section~\ref{sec:con}.

\section{Methods} \label{sec:methods}

In this section, we briefly recall the RI data model in the context of wide-field imaging and provide a summary of AIRI, building from the underlying theory \citep{Terris22} and its first application to real RI data \citep{dabbech2022first}. We also outline the encompassing framework for wide-field imaging, focusing on the parallelisation of the AI denoiser and the automated selection of associated parameters. A description of the framework's underpinning wide-field parallel measurement operator can be found in Section 2 of Part I.

\subsection{RI data model}
We recall the discrete RI data model in the context of wide-field imaging, detailed in Part I, whereby the measured visibilities $\yb \in \eC^M$ are modelled from a discrete representation of the sought radio image $\xb \in \eR_+^N$ as follows
\begin{equation}
    \label{eq:datamodel}
    \yb = \bm\Phi \xb + \nb,
\end{equation}
where $\nb \in \eC^M$ is a realisation of random Gaussian noise with mean zero and standard deviation $\tau>0$, and $\bm\Phi \in \eC^{M \times N}$ is the measurement operator encompassing the Fourier sampling and the so-called $w$-effect, a chirp-like phase modulation emanating from the non-coplanarity of the array \citep{Cornwell2008}. Often, a noise-whitening operation is applied to the measured visibilities to ensure constant standard deviation of the noise \citep[see Appendix A of ][for more details]{Terris22}. The operation can be applied in combination with a weighting scheme compensating for the highly non-uniform Fourier sampling \citep[\emph{e.g.} Briggs weighting; ][]{briggs95} to enhance the effective resolution of the observation. Naturally, any transform applied to the data is injected in the model of the measurement operator.
Our imaging framework is shipped with a  parallel and memory-efficient measurement operator, ensuring scalability to large data sizes \citep[see][ for a comprehensive summary]{dabbech2022first}. 

\subsection{AIRI algorithm}

The recent PnP scheme established that proximal optimisation algorithms, such as FB, enable not only the use of proximal operators of handcrafted regularisation operators, but also the injection of learned DNN denoisers, which define regularisation implicitly \citep{venkatakrishnan2013,romano2017}. In order to preserve the convergence of the algorithm, and the interpretability of its solution, the PnP denoiser must typically satisfy a “firm non-expansiveness” constraint, ensuring that it contracts distances \citep{pesquet2020learning,hurault2022proximal}. Learning denoisers from rich databases (as opposed to handcrafting proximal operators) opens the door to more powerful regularisation. The speed of DNNs on GPU also offers a significant acceleration over iterative proximal operators. In this context, the AIRI imaging algorithm \citep{Terris22} is underpinned by the same FB structure as the uSARA imaging algorithm (see Section 2 in Part I) with the image update at each iteration alternating between a `forward' step enforcing data fidelity with respect to \eqref{eq:datamodel}, and a `backward' denoising step for image regularisation:
\begin{equation} \label{eq:dnn}
 (\forall k\in \eN) \qquad \xb^{(k+1)} = {{\rm D}}\left(  \xb^{(k)} -\gamma \nabla f (\xb^{(k)}) \right).
\end{equation}
The operator ${{\rm D}}$ denotes a learned DNN denoiser. Considering the standard data-fidelity function $f(\xb; \yb) = 1/2 \|\yb -\bm\Phi \xb\|^2_2$, where $\|.\|_2$ denotes the $\ell_2$-norm of its vector argument, the operator $\nabla f $ stands for its the gradient and reads $\nabla f (\xb) = \text{Re}\{\bm{\Phi}^\dagger\bm{\Phi}\} \xb -\text{Re}\{\bm{\Phi}^\dagger\yb\}$. The parameter $\gamma>0$ is a sufficiently small step size.

\subsection{DNN training and noise level}
\label{ssec:dnnselection}

Following \citet{Terris22}, the AIRI denoisers used in this work were trained in a supervised approach to remove random Gaussian noise with zero-mean and standard deviation $\widehat{\sigma}>0$ from noisy input images. The denoisers rely on a simple denoising convolutional neural network (DnCNN) architecture and are trained using a rich high-dynamic range database synthesised from  optical astronomy images, with groundtruth images normalised to have a peak value equal to 1. Importantly, the training loss function is regularised with an appropriate non-expansiveness term on the denoiser ${{\rm D}}$.

Given the normalisation of the groundtruth images from the training database, the DNN's noise level can be interpreted as the inverse of a target dynamic range, which can intuitively be adjusted to match the signal-to-noise ratio in the observed data. Considering $L$ as the spectral norm of $\text{Re}\{\bm{\Phi}^\dagger \bm{\Phi}\}$, the standard deviation of the measurement noise in the image domain can be estimated as $\sigma=\eta\tau/\sqrt{2L}$ \citep{thouvenin22,wilber221} where $\eta>0$ is derived from the data-weighting operator when considered in imaging and is set to 1 otherwise. Hence, the target dynamic range of the sought image is given by $\operatorname{max}_j\{{x}_j\}/\sigma$. 
However, the peak value of the true image of the sky is not accessible in practice. We therefore resort to the dirty image defined as $\overline{\xb}^{\textrm{dirty}}=\beta \text{Re}\{\bm{\Phi}^\dagger \yb \}\in\eR^N$, where $\beta>0$ is a normalisation factor\footnote{The factor $\beta$ corresponds to the peak value of the non-normalised point spread function given by $\text{Re}\{\bm{\Phi}^\dagger \bm{\Phi}\}\bm{\delta}$, where $\bm{\delta} \in \eR^N$ is the image with one at its centre and zero otherwise.}, and approximate the peak value of the sought image by the peak value of the dirty image $\kappa=\operatorname{max}_j\{{\overline x}^{\textrm{dirty}}_j\} >0$. The value of $\kappa$ constitutes an upper bound on the estimated peak value\footnote{In our experiments, we observed that $\kappa$ is within one order of magnitude from the peak value of the reconstructed image.}. Consequently, $\kappa/\sigma$ provides an estimate (in fact, an upper bound) on the target dynamic range.

In this context, the RI inverse problem \eqref{eq:datamodel} is normalised by $\kappa$ to ensure that the sought image $\xb/\kappa$ satisfies the same normalisation constraints of the training images, such that pixel values are below 1. The DNN denoiser is trained for the removal of a zero-mean random Gaussian noise with standard deviation
\begin{equation}
\label{eq:heuristic}
    \widehat{\sigma}=\sigma/{\kappa} \textrm{, where~} \sigma=\eta\tau/\sqrt{2L}.
\end{equation}
The AIRI image estimate is later re-scaled back via multiplication by $\kappa$. 

The apparent dependency of the training noise level on the statistics of the noise corrupting the RI data and the associated measurement operator raises a generalisability concern for the AIRI denoisers. However, this can be  circumvented via further scaling of the inverse problem to bring the target dynamic range of the reconstruction to the inverse of the noise level of an already available DNN denoiser. 

\subsection{Denoiser selection} \label{ssec:denoise-strat}
The selection of the appropriate denoiser can be conducted via two approaches. The first approach relies on a pre-trained \emph{shelf of denoisers}, from which the appropriate denoiser is selected depending on the target dynamic range of the reconstruction. A set of denoisers are thus trained, with noise levels sampled within a wide range of values, reflecting a whole range of dynamic ranges of interest in modern RI imaging. For each RI dataset, AIRI's denoiser is selected from the shelf as the DNN with the nearest noise level ${\sigma}_s$ below the inverse of the target dynamic range $\widehat{\sigma}$. This implies considering a slightly looser image peak upper bound $\kappa\widehat{\sigma}/{\sigma}_{s}$ for the re-scaling of the inverse problem \eqref{eq:datamodel}, thus leading to a heuristic value ${\sigma}_{s}$ for the training noise level in \eqref{eq:heuristic}. 

The second approach leverages a pre-trained \emph{single (universal) denoiser} to be applied for the image formation of any RI dataset. The denoiser is trained with a very low noise level ${\sigma}_u$, tailored for the highest target dynamic ranges of interest for modern RI imaging. For any RI dataset of interest, this amounts to considering a possibly much looser image peak upper bound $\kappa\widehat{\sigma}/{\sigma}_{u}$ for the re-scaling of the inverse problem \eqref{eq:datamodel}, systematically leading to a heuristic value ${\sigma}_{u}$ for the training noise level in \eqref{eq:heuristic}. This second approach was already shown to be efficient in the formation of high-quality radio maps when applied to observations from the MeerKAT telescope \citep{dabbech2022first}.

\subsection{Denoiser faceting}
Owing to their convolutional nature and narrow receptive fields, AIRI's denoisers can be applied to facets of the image without causing any faceting-related artefacts, provided that appropriate facet overlaps are considered. This feature enables the scalability of AIRI denoisers to large image dimensions through their parallel application to image facets, as well as circumventing the memory limitation of the GPUs during inference.

 \begin{figure}
 \centering
 \includegraphics[width=.49\textwidth]{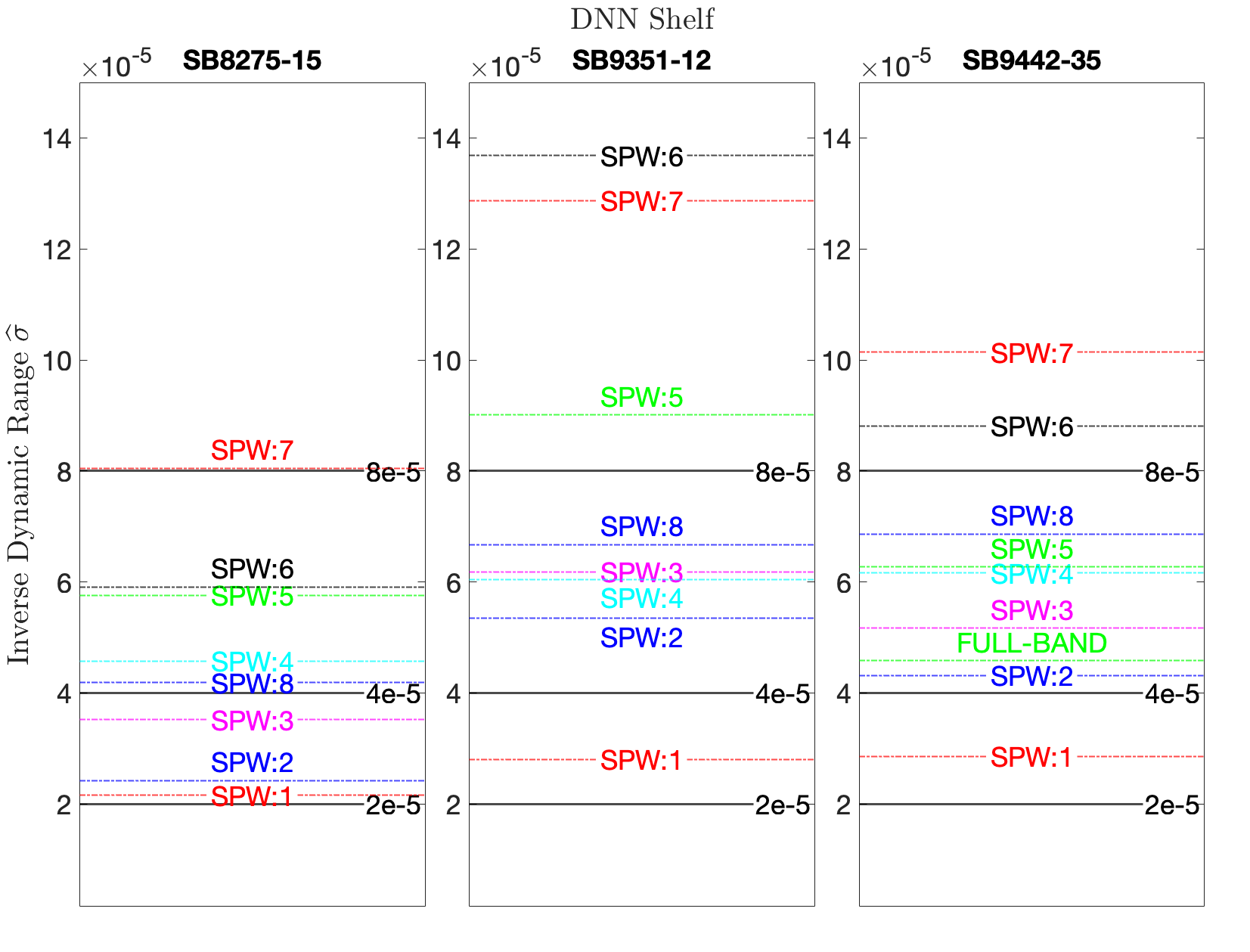}
 \caption{The positioning of the inverse of the target dynamic range, $\widehat{\sigma}$, associated with the different SPWs of the selected fields (dashed lines), with respect to the noise level, $\sigma_s$, of the pre-trained shelf of DNN denoisers (black solid lines). From left to right: plots for the fields SB8275-15, SB9351-12, and SB9442-35, respectively. Specifically to SB9442-35, we also add the inverse of the target dynamic range associated with the full-band imaging experiment. 
\label{DNNshelf}}
 \end{figure}

\section{Data, imaging, and analysis} \label{sec:data}

In validating the AIRI algorithm, we consider the same RI datasets imaged in Part I. These data consist of three individual beam observations from ASKAP Early Science and Evolutionary Map of the Universe Pilot survey \citep[EMU-PS][]{2021PASA...38...46N} scheduling blocks (SBs): SB8275-15, SB9351-12, and SB9442-35. The measurement sets containing calibrated visibilities of these selected observations were produced by ASKAPsoft \citep{2021PASA...38....9H} and obtained from the CSIRO ASKAP Science Data Archive (CASDA; \citealp{2017ASPC..512...73C}. For our imaging purposes, each of these observations, with bandwidths of 288~MHz (ranging between [800,1158]~MHz), was split into eight spectral windows (SPWs). For the first two fields (SB8275-15 and SB9351-12), the resulting sub-band data were imaged separately to form monochromatic images with dimensions of $5500 \times 5500$ pixels and a cell-size of 2.2~arcsec. For the third field (SB9442-35), sub-band images and a single full-band image with dimensions of $4096 \times 4096$ pixels and cell-sizes of 2.2~arcsec were formed. The full-band data of SB9442-35 were reconstructed into a monochromatic full-band image with the aim to increase both the dimensionality and the sensitivity of the data. Full details of the data and the imaging settings can be found in Tables~1 \& 2 of Part I.

Specific to AIRI, in each imaging experiment we enabled the image-faceting functionality of the denoiser by splitting the image into four facets of equal dimensions. Faceting was found to be necessary for satisfying memory requirements. In the selection of the appropriate denoiser, we first determined the values of the inverse of the target dynamic range, $\widehat{\sigma}$, of all formed sub-band and full-band images, following \eqref{eq:heuristic}. We primarily opted for the pre-trained shelf strategy for all AIRI reconstructions, where the considered noise levels of the pre-trained denoisers are $\sigma_s = [2,4,8]\times 10^{-5}$. 
We also investigated the pre-trained universal denoiser strategy for the field SB9351-12. In this case, the universal denoiser is chosen from the pre-trained shelf as the denoiser with the lowest noise level (equivalently, the highest dynamic range), that is $\sigma_u = 2\times10^{-5}$.

We note that training under the firm non-expansiveness constraint is highly challenging. While \citet{Terris22} demonstrated in simulation that the AIRI training approach leads to a robust way to ensure convergence of the PnP algorithms, we acknowledge that, when used for real data and at large image sizes and dynamic ranges such as those of interest here, some denoisers lead to algorithm instability, ultimately requiring further training. This phenomenon was also witnessed by \citet{dabbech2022first}.

AIRI experiments were run on Cirrus\footnote{\url{http://www.cirrus.ac.uk}}, a UK Tier2 high-performance computing (HPC) service, and utilised its GPU compute nodes. A Cirrus GPU node is comprised of 4 GPUs and 40 CPU cores with 384~GB of shared memory. 
Each imaging experiment of AIRI is launched on one to five GPU nodes, depending on the memory requirements,  whereby CPU cores, allocated dynamically, were utilised for the forward step, more precisely the application of the measurement operator, and GPUs were exploited in the parallel application of the denoising DNN on facets of the image.

For a quantitative assessment of AIRI's performance, we focus on the same primary sources of interest as presented in Part I and analyse their associated flux measurements and spectral index maps. Results are compared to those obtained with uSARA and {\tt WSClean} for all spectral windows of each imaged field. The AIRI flux measurements were computed from hand-drawn regions (generated using the visualisation software SAOImageDS9; \citealp{ds903}) slightly different from those considered in the uSARA and {\tt WSClean} images, to better match the morphology of recovered emission. Spectral index maps inferred from the AIRI-ASKAP sub-band images were also obtained following the same procedure described in Part I. Likewise, only the first six sub-band images were used to generate the spectral index maps due to the limited diffuse signal recovered in the final two sub-bands. Similarly to uSARA, blurring with a circular Gaussian beam of 5 arcsec was applied to AIRI-ASKAP sub-band images to smooth the source structure before flux measurements were fitted to the spectral curve: $S_{\nu} \propto \nu^{-\alpha}$, where $S_{\nu}$ is the flux density for a given beam area at a given frequency $\nu$ and $\alpha > 0$ is the spectral index. In presenting our AIRI images, we also make use of optical images from the first data release of the Dark Energy Survey (DES; \citealp{2018ApJS..239...18A}).

We performed an additional assessment towards the quantification of epistemic model uncertainty by comparing variations of AIRI (\emph{i.e.} using different DNN denoisers) when imaging each spectral window of the field SB9351-12. More details on this analysis can be found in Section~\ref{ssec:results-strat}.

\begin{figure*}
\centering
\includegraphics[width=\textwidth]{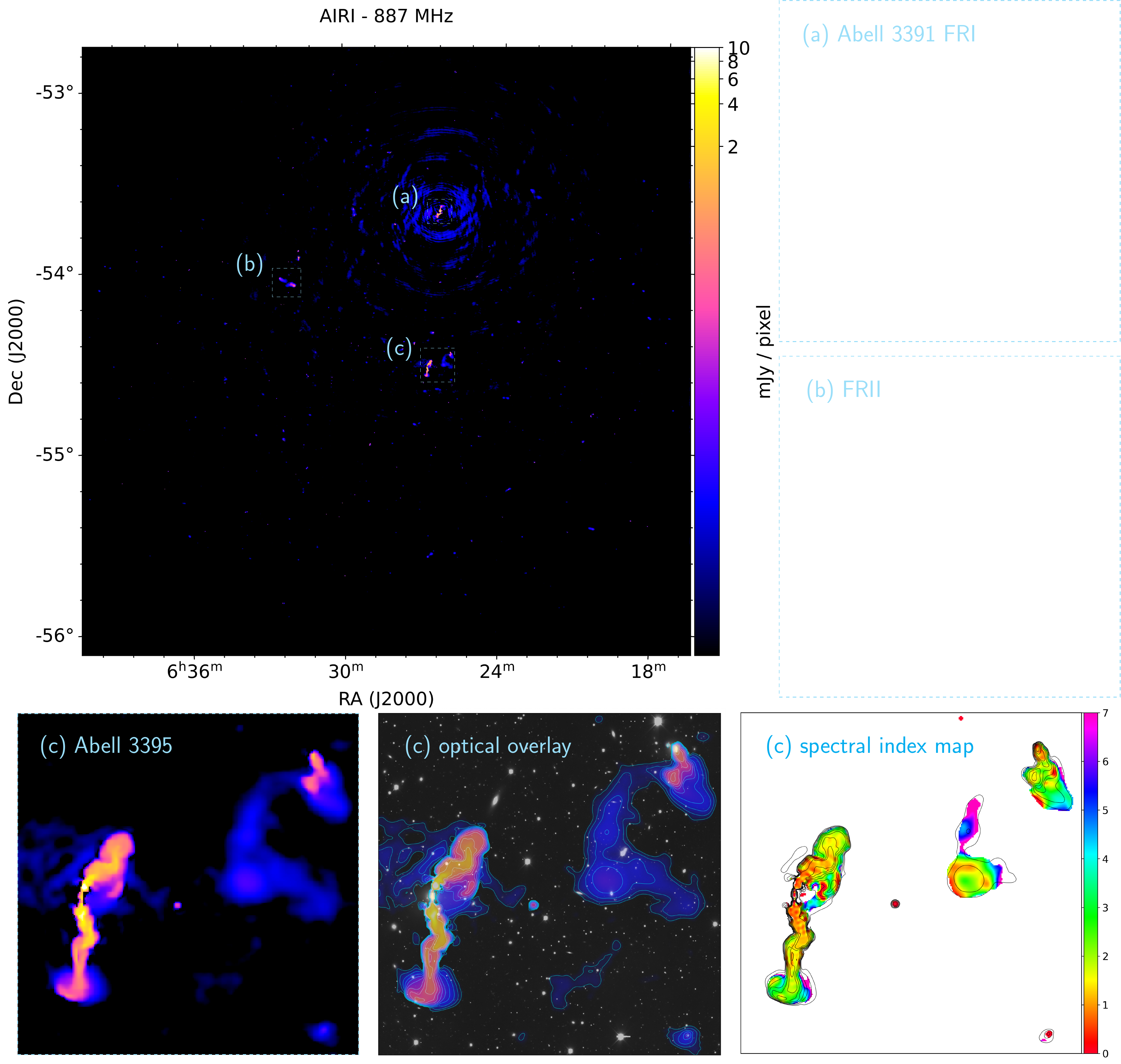}
\caption{
SB8275-15 -- AIRI: Full FoV image covering the merging cluster system Abell 3391-95, at the first sub-band (SPW:1, centred at 887 MHz). For visual comparison with {\tt WSClean} and uSARA, refer to their respective Figures 1 and 2 from Part I. This monochromatic image is an AIRI model with a pixel resolution of $2.2 \times 2.2$ arcsec. Panel (a) centred on the FR I radio galaxy in A3391; panel (b) centred on cluster member FR II radio galaxy; (c) panels centred on FR I and diffuse source in A3395. Middle (c) panel: r-band optical image from DES overlaid with AIRI model image, demarcated by blue contours at levels $\{2^n\}_{0 \leq n \leq 10} ~\mu$Jy pixel$^{-1}$. Rightmost (c) panel: spectral index map obtained with the first six sub-band images of AIRI after smoothing with a common circular Gaussian beam of 5 arcsec. In \citet{askapdataset} are provided all sub-band images combined into the GIF \texttt{`SB8275-15\_AIRI'}, and the spectral index map of Abell 3395 obtained with AIRI, uSARA, and {\tt WSClean} in the GIF \texttt{`SpectralIndexMap\_Abell\_$3395$'}, together with a colour blind-friendly version in the GIF  \texttt{`SpectralIndexMap\_Abell\_$3395$\_colorblind\_friendly'}.
 }
\label{A3391airi}
\end{figure*}

\begin{table*}
\begin{tabular}{| *{9}{c|} }
\hline
\hline
\textbf{A3395 Phoenix} & $S_{887}$  & $S_{923}$  & $S_{959}$ & $S_{995}$  & $S_{1031}$  & $S_{1067}$  & $S_{1103}$ & $S_{1139}$  \\
    \hline
AIRI model  &   25.3 & 27.0 & 17.9 & 8.5 & 11.3 & 12.1 & 9.7 & 1.6  \\ 
\hline

\end{tabular}
\caption{Integrated flux density values in [mJy] of the diffuse phoenix source in Abell 3395 for each SPW imaged with AIRI. Central frequency of each SPW is listed in MHz. See Table 3 in Part I for uSARA and {\tt WSClean} flux measurements of the diffuse phoenix source. \label{tab:fluxA3391}}
\end{table*}


\section{Results} \label{sec:results}

In this section, we showcase high-resolution high-fidelity images of our three selected fields produced by the AIRI algorithm encapsulated in our parallelised and automated imaging framework and investigate two approaches for the selection of the DNN denoiser, as described in Section~\ref{ssec:dnnselection}.   
Figure~\ref{DNNshelf} illustrates the positioning of the inverse of the target dynamic range $\widehat{\sigma}$ of each imaged spectral window for each field with respect to the noise levels of the learned DNNs. 

\begin{figure*}
\centering
\includegraphics[width=\textwidth]{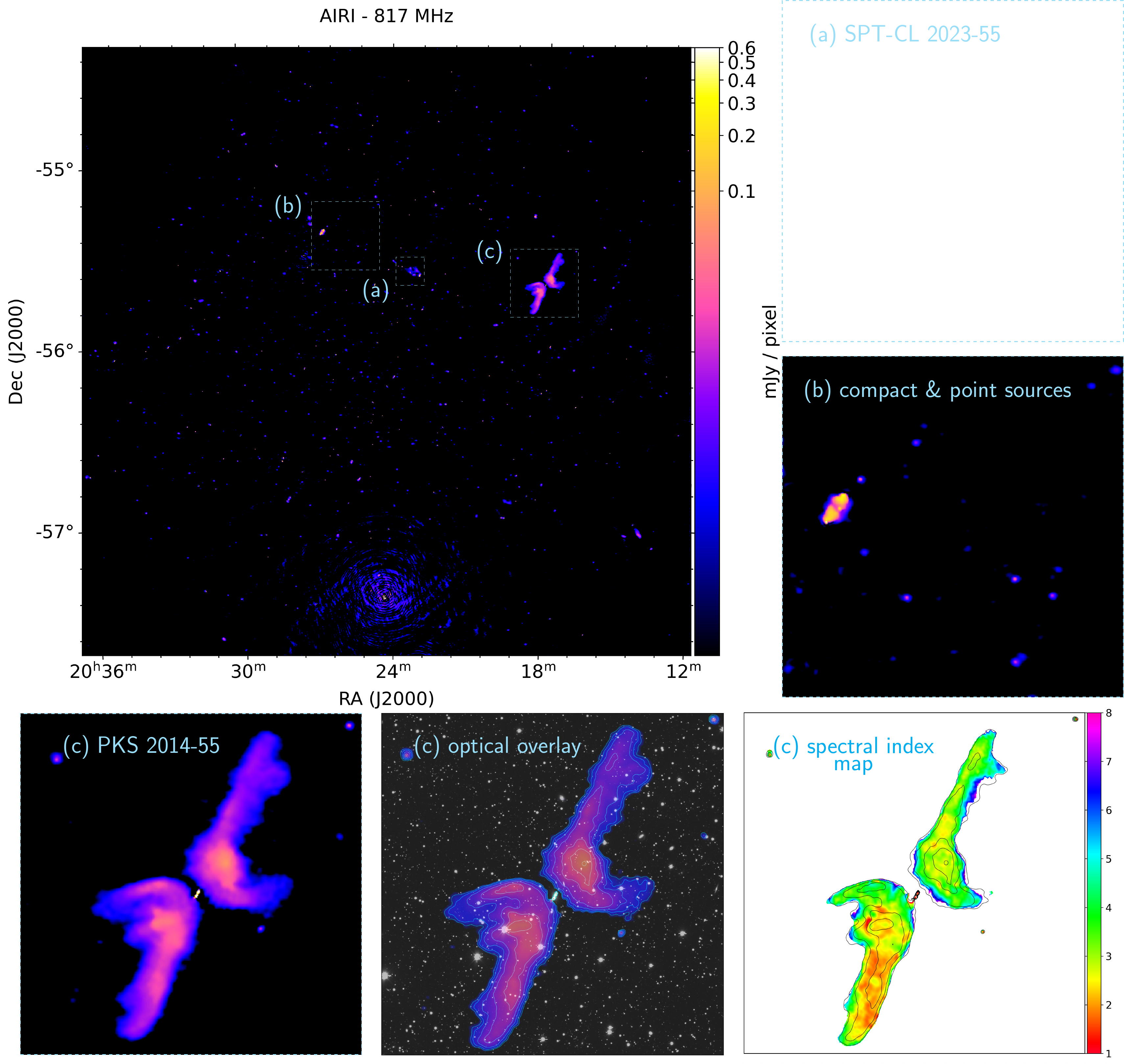}
\caption{SB9351-12 -- AIRI: Full FoV image covering the merging cluster SPT2023 and the X-shaped radio galaxy PKS 2014-55, at the first sub-band (SPW:1, centred at 817 MHz). For visual comparison with {\tt WSClean} and uSARA, refer to their respective Figures 3 and 4 from Part I. This monochromatic image is an AIRI model with a pixel resolution of $2.2 \times 2.2$ arcsec. Panel (a) centred on the merging galaxy cluster SPT2023; panel (b) centred on a field containing compact and point sources; (c) panels centred on the X-shaped radio galaxy PKS 2014-55. Middle (c) panel: r-band optical image from DES overlaid with the AIRI model image, demarcated by blue contours at the levels $\{2^n\}_{0 \leq n \leq 10} ~\mu$Jy pixel$^{-1}$. Rightmost (c) panel: spectral index map obtained with the first six sub-band images of AIRI after smoothing with a common circular Gaussian beam of 5 arcsec. In \citet{askapdataset} are provided all sub-band images combined into the GIF \texttt{`SB9351-12\_AIRI'}, and the spectral index map of the X-shaped radio galaxy obtained with AIRI, uSARA, and {\tt WSClean} in the GIF \texttt{`SpectralIndexMap\_PKS\_2014\_55'}, together with a colour blind-friendly version in the GIF  \texttt{`SpectralIndexMap\_PKS\_2014\_55\_colorblind\_friendly'}.
\label{SPT2023airi}}
\end{figure*}

Select AIRI-ASKAP images are displayed in Figures~\ref{A3391airi}--\ref{9442airi}, in a format identical to the uSARA-ASKAP and {\tt WSClean} images presented in Part I. The AIRI-ASKAP figures consist of full FoV images with zoomed-in views focused on complex radio emission of interest, and their associated optical images and spectral index maps. In \citet{askapdataset}, we provide all sub-band AIRI-ASKAP images of the three selected fields (as FITS files) and combine them into GIF files to better show how emission and source morphology changes over the full frequency band. Throughout this section, we refer the reader to Part I for specific comparisons of our imaging results with the uSARA and {\tt WSClean} reconstructions from Part I. 

Upon visual inspection, our monochromatic AIRI-ASKAP images of all three fields capture more extended structure than seen in the pure-optimisation counterpart uSARA-ASKAP images. In particular, the faintest structures of our diffuse targets of interest appear more pronounced and defined than they did in the uSARA-ASKAP images. However, the intensity of the faintest point sources, as reconstructed by AIRI, seems to be diminished. In the following subsections, we focus on each ASKAP field and address these differences by examining specific sources of interest. We present detailed comparisons of source morphology, flux density measurements, and spectral index maps between the three imaging algorithms. In an experiment toward uncertainty quantification of AIRI denoisers, we also showcase AIRI reconstructions made via the universal denoiser approach.

\subsection{First field: SB8275-15}
This field contains the massive, merging galaxy clusters Abell 3391 (in the north) and Abell 3395 (in the south). The cluster pair is connected by a warm gas bridge, recently discovered in eROSITA X-ray observations \citep{2021A&A...647A...2R}. In Figure~\ref{A3391airi}, we present our AIRI image of this full imaged FoV (3.36$^{\circ}$) of the first spectral window (SPW:1). The figure includes zoomed-in views of the FR-I in Abell 3391 (a: top right panel), a FR-II cluster member in the east (b: middle right panel), and multiple sources in Abell 3395 (c: bottom panels). The bent-tail FR I radio galaxies at the centres of Abell 3391 and Abell 3395 (see Table 1 in Part I for source names) are reconstructed with similar brightness and resolution when compared to our uSARA image from Part I (the peak pixel flux of the FRI in Abell 3391 is 20~mJy in both the AIRI and uSARA images). The highly resolved detail of the `braiding' in these FRI jets, not resolved in {\tt WSClean} images, is present in both the AIRI and uSARA images. However, the radio galaxies as captured by AIRI exhibit more blended edges than seen in the uSARA image. In addition, the ring-like artefacts emanating from these bright FRI sources take on a much more extended structure in our AIRI image -- their appearance is dimmed and they propagate further out from their point of origin. Mainly, there is a noticeable difference in the diffuse structure recovered in the candidate phoenix of Abell 3395 and the background FR-II radio galaxy (c and b panels, respectively, in Figure~\ref{A3391airi}).


\begin{table*}
\begin{tabular}{| *{9}{c|} }
    \hline
    \hline
    \textbf{X-Shaped RG} & $S_{817~{\rm MHz}}$  & $S_{853~{\rm MHz}}$  & $S_{889~{\rm MHz}}$ & $S_{925~{\rm MHz}}$  & $S_{961~{\rm MHz}}$  & $S_{997~{\rm MHz}}$  & $S_{1033~{\rm MHz}}$ & $S_{1069~{\rm MHz}}$  \\
    \hline
    \hline
AIRI model   &   678.7 & 580.0 & 485.0 & 427.7 & 436.2 & 352.7 & 302.3 & 190.5 \\
    \hline

\end{tabular}
\caption{Integrated flux density values in [mJy] of the X-shaped radio galaxy PKS 2014-55 for each SPW imaged with AIRI. The listed flux densities are totals from summing the flux densities measured in regions mapping the east wing, the west wing, and the core. See Table 4 in Part I for uSARA and {\tt WSClean} flux measurements of the X-shaped radio galaxy.\label{tab:flux-xshape}}
\end{table*}

\begin{table*}
\begin{tabular}{| *{9}{c|} }
    \hline
    \hline
    \textbf{SPT2023 Relic} & $S_{817~{\rm MHz}}$  & $S_{853~{\rm MHz}}$  & $S_{889~{\rm MHz}}$ & $S_{925~{\rm MHz}}$  & $S_{961~{\rm MHz}}$  & $S_{997~{\rm MHz}}$  & $S_{1033~{\rm MHz}}$ & $S_{1069~{\rm MHz}}$  \\
    \hline
AIRI model   &  4.3 & 3.2 & 1.8 & 2.0 & 2.8 & 2.3 & 1.7 & 0.5 \\
    \hline
\end{tabular}
\caption{Integrated flux density values in [mJy] of the radio relic in SPT2023 for each SPW imaged with AIRI. See Table 5 in Part I for uSARA and {\tt WSClean} flux measurements. \label{tab:flux-SPT2023}}
\end{table*}


\subsubsection{Abell 3395}
In the middle panel (c) of Figure~\ref{A3391airi}, contours mapping the AIRI recovered emission above $1~\mu$Jy pixel$^{-1}$ are overlaid on an r-band optical image from DES. In comparison with uSARA, the recovered emission of the candidate phoenix by AIRI seems to grow in size. Although appearing slightly fainter and smoother, more extended structure is revealed in the AIRI reconstruction of the Abell 3395 phoenix, as the north-west arm now bridges the dim core and the compact sources at the north-west edge of the cluster. The structure of the recovered emission also changes from one spectral window to the next, with noticeable fading as the frequency increases \citep[see associated GIF in ][]{askapdataset}. Flux density measurements of the phoenix from the sub-band AIRI-ASKAP images are provided in Table~\ref{tab:fluxA3391}. In comparing the flux density measurements to those taken from uSARA and {\tt WSClean} images, we see some slight variations across the spectral windows with the exception of the highest frequency where there is clearly less flux recovered in the AIRI image.

The most exciting result is perhaps the improvement of the spectral index map of the phoenix obtained with AIRI when compared to uSARA (see rightmost panel (c) of Figure~\ref{A3391airi}). Since more diffuse structure is recovered by AIRI overall, even at the higher frequencies, the spectral index map has more coverage over the full source morphology. This coverage aids in source classification, enabling the identification of a trend in the spectral index as the emission shifts from the dim core to the north-west and south-west arms. There is clearly more steepness of the spectra as compared to the uSARA map shown in Part I. The dim core recovered by AIRI has a steeper index ($2.1 \leq \alpha \leq 2.8$), and the north-west arm shows a sharp, rather than gradual, drop-off from the core. There still exists the ring of flatter emission around the core, matching the results from uSARA and {\tt WSClean}. Our AIRI results are in line with the hypothesis that this source is no longer receiving a fresh injection from an active nucleus and that the surrounding emission may be undergoing some gentle re-energisation, which in turn is causing brightening and flattening of old and faded AGN emission. Interestingly, both the FR-I to the east and the compact source to the north-west exhibit consistent spectral behaviour between uSARA and AIRI.

\subsection{Second field: SB9351-12}
\label{ssec:SB9351}
The second selected field  covers the merging galaxy cluster SPT-CL J2023-5535 (hereafter SPT2023) and the X-shaped radio galaxy PKS 2014-55. As stated in Part I, two recent studies have been separately published for these sources of interest: \citet{2020ApJ...900..127H} confirmed the detection of a radio halo and radio relic in SPT2023 with the same data used in this work and \citet{2020MNRAS.495.1271C} used MeerKAT observations to generate total intensity, polarisation, B-field maps, and spectral index maps of the X-shaped radio galaxy. In Figure~\ref{SPT2023airi}, we present our AIRI image of the full FoV (3.36$^{\circ}$) of the first spectral window (SPW:1) of SB89351-12. The figure includes zoomed-in views on the merging cluster SPT2023 (a: upper right panel), a field of compact and point-like sources (b: middle right panel), and the X-shaped radio galaxy PKS 2014-55 (c: bottom panels). 

In comparison with the uSARA image (Figure 3 of Part I), there is an undeniable improvement in the recovery of faint emission within the zoomed-in views seen in the AIRI image. The diffuse emission stretching east-to-west in the {\tt WSClean} image of SPT2023, which was not recovered by uSARA, clearly emerges in the AIRI image. Similarly, faint point sources seen in the {\tt WSClean} image but not in the uSARA image, are captured by AIRI, though appearing to be somewhat fainter and smoother (see the panel (b) of Figure~\ref{SPT2023airi}). Finally, the calibration artefacts are still noticeable at the southern edge of the pointing, taking the form of ring-type artefacts emanating from the bright quasar RX J2024.3-5723 and propagating radially up to 1 deg. Compared to uSARA, these artefacts are generally fainter but extend further. 


\begin{figure*}
\centering
\includegraphics[width=\textwidth]{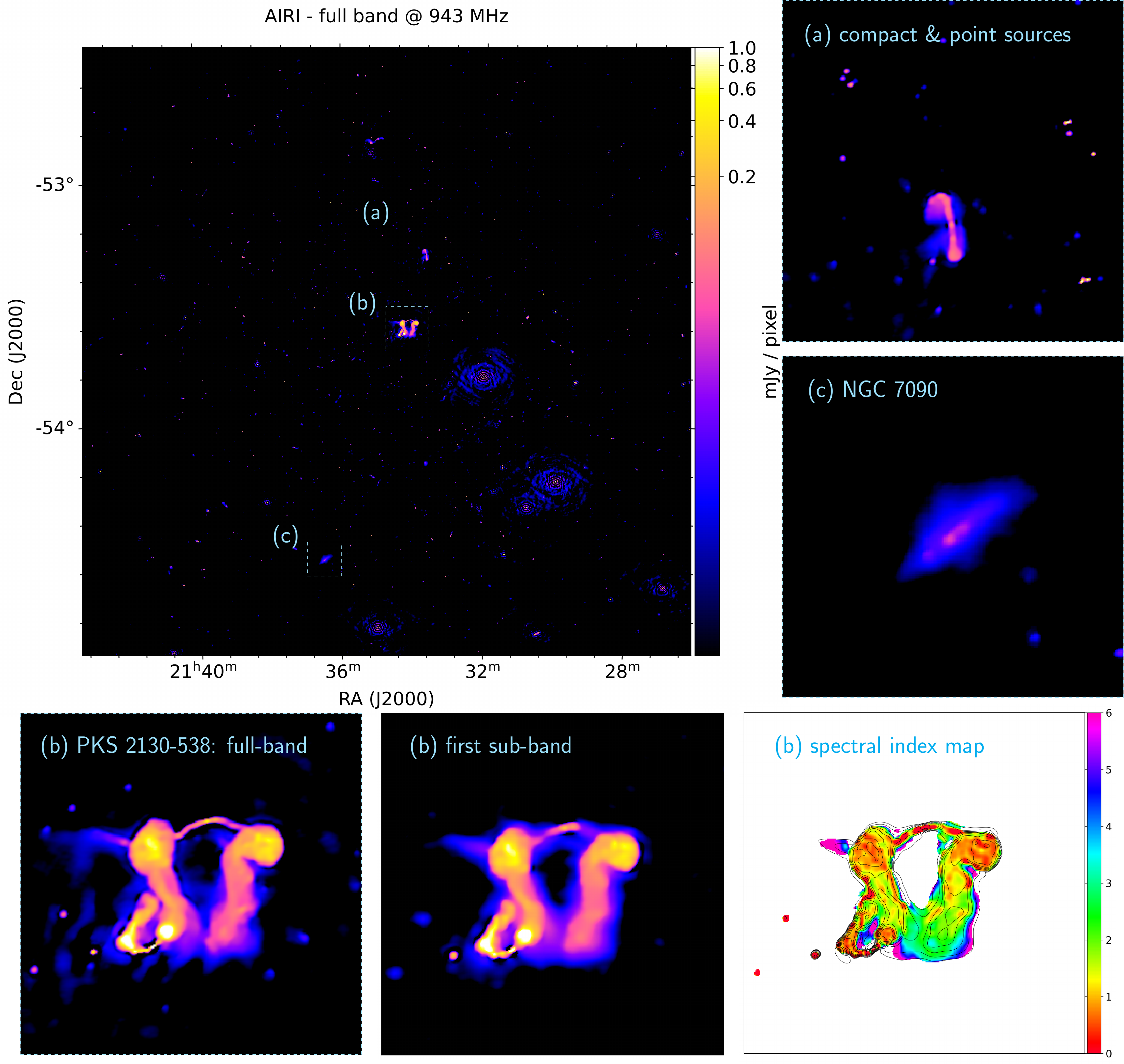}
\caption{SB9442-35 -- AIRI: Full FoV image covering PKS 2130-538, formed using the full-band data (centred at 943 MHz). For visual comparison with {\tt WSClean} and uSARA, refer to their respective Figures 5 and 6 from Part I. This monochromatic image is an AIRI model with a pixel resolution of $2.2 \times2.2$ arcsec. Panel (a) centred on a field containing extended and point-like radio galaxies; panel (c) centred on the star-forming galaxy NGC 7090; (b) panels centred on ``the dancing ghosts'' (PKS 2130-538). Middle (b) panel: image made with only the first sub-band of data (SPW:1, centered at 817), shown for a comparison of sensitivity. Rightmost (b) panel: spectral index map made with the first six sub-band images of AIRI after smoothing with a common circular Gaussian beam of 5 arcsec. In \citet{askapdataset} are provided all sub-band images combined into the GIF \texttt{`SB9442-35\_AIRI'}, and the spectral index map of ``the dancing ghosts'' obtained with AIRI, uSARA, and {\tt WSClean} in the GIF \texttt{`SpectralIndexMap\_PKS\_2130\_538'}, together with a colour blind-friendly version in the GIF  \texttt{`SpectralIndexMap\_PKS\_2130\_538\_colorblind\_friendly'}.
\label{9442airi}
}
\end{figure*}

\begin{table*}
\begin{tabular}{| *{10}{c|} }
    \hline
    \hline
    \textbf{Dancing Ghosts} & $S_{\rm fullband - 943~{\rm MHz}}$ & $S_{817~{\rm MHz}}$  & $S_{853~{\rm MHz}}$  & $S_{889~{\rm MHz}}$ & $S_{925~{\rm MHz}}$  & $S_{961~{\rm MHz}}$  & $S_{997~{\rm MHz}}$  & $S_{1033~{\rm MHz}}$ & $S_{1069~{\rm MHz}}$  \\
    \hline
AIRI model  &  116.7 & 128.3 & 123.8 & 118.1 & 113.2 & 109.2 & 105.7 & 102.6 & 97.7 \\
    \hline

\end{tabular}
\caption{Integrated flux density values of ``the dancing ghosts'' PKS 2130-53 for each SPW imaged with AIRI. See Table 6 in Part I for uSARA and {\tt WSClean} flux measurements of ``the dancing ghosts''. \label{tab:fluxghosts}}
\end{table*}


\subsubsection{X-shaped Radio Galaxy}
In the middle panel (c) of Figure~\ref{SPT2023airi}, we overlay AIRI-ASKAP emission of the X-shaped radio galaxy as contours on an r-band optical map from DES. Compared to uSARA, the X-shape radio galaxy as reconstructed by AIRI appears to have a greater extent of diffuse emission, though with a noticeable loss in the resolution of the compact structure within the lobes. This behaviour is consistent across all the sub-band images of AIRI, where smoother and more diffuse edges are observed. Table~\ref{tab:flux-xshape} reports the measured flux densities per spectral window for the X-shaped radio galaxy. AIRI flux measurements are consistently lower than the uSARA flux measurements for SPW:1 through SPW:5, then increase for SPW:6 and SPW:7, and drop lower again in spectral window 8.

A spectral index map of the X-shaped radio galaxy is shown in the rightmost (c) panel of Figure~\ref{SPT2023airi}. There is an incredible improvement in the coverage of the AIRI spectral index map compared to uSARA, thanks to the diffuse flux consistently recovered at the edges of the lobes, even at the higher frequencies, with AIRI. The -- most likely -- artificial steepening at the edges of the lobes seen in the uSARA spectral index map is now corrected in the AIRI map. Nonetheless, AIRI recovers slightly steeper spectra on the borders of the lobes when compared to the {\tt WSClean} spectral index map. Owing to the superb resolution achieved by AIRI, turbulent activity can be traced where plasma in the lobes exhibits a flatter spectral index. The south-east leg of the east wing shows a spectral index of $1.4 \leq \alpha \leq 2.1$, which is much flatter than the emission in the west wing (with an average spectral index of about $\alpha = 3$. The furthest north-west portion of the west wing exhibits an ultra-steep spectrum $3.5 \leq \alpha \leq 5$. Wide-band deconvolution is necessary to confirm these ultra-steep values. 

\subsubsection{SPT-CL J2023-5535}
In Part I, our monochromatic {\tt WSClean} image shows the radio halo in SPT2023 as an increase in noise at the cluster centre, but our uSARA image did not show any diffuse structure resembling a radio halo. In Figure~\ref{SPT2023airi}, the panel (a) focuses on the diffuse emission present in SPT2023. Here, our AIRI image does in fact recover the diffuse structure of the radio halo -- elongating from the western relic towards the east. Across the sub-bands the radio halo is most clearly detected in AIRI images SPW:1 and SPW:2. It is quite remarkable that the radio halo is detected in these narrow, sub-band AIRI images since the full 288~MHz bandwidth (in the form of a wideband {\tt WSClean} image) was used to detect and measure the radio halo's signal in \citet{2020ApJ...900..127H}. We also find that the SPT2023 radio relic has a smoother, wider, and fainter morphology when compared to uSARA. AIRI flux measurements of the relic for SPW:6 and 7 are slightly higher than uSARA flux measurements, but lower for all other spectral windows, as reported in Table~\ref{tab:flux-SPT2023}.

\subsection{Third field: SB9442-35}

This final selected field is centred on the complex radio source PKS 2130-538, nicknamed ``the dancing ghosts,'' owing to its peculiar and mirrored ghost-like shape. Two radio lobes, bridged by arching jets from the primary AGN host, extend southwards and blend into each other. A secondary AGN in the south-east produces a similar arched jet with a bent-tail that curls back around to the eastern primary lobe. With the original images generated from the ASKAP Evolutionary Map of the Universe Survey (EMU; \citealp{2011PASA...28..215N}), \citet{2021PASA...38...46N} mention that the interaction between the primary and secondary AGN is unclear. With our super-resolved uSARA images, presented in Part I, we have been able to distinguish a clear physical separation between the secondary curling jet and the primary western lobe. Nonetheless, this strange source offers an interesting case study of turbulent dynamics in bent-tail radio galaxies. 

For this field, we produced eight sub-band images as well as a monochromatic full-band image with AIRI. In Figure~\ref{9442airi}, we present the AIRI image of a $\sim 2.5^{\circ}$ FoV formed using the full-band (288 MHz) data of SB89442-35. This figure includes zoomed-in views on a field containing an extended radio galaxy and points sources (a: upper right panel), the star-forming galaxy NGC 7090 (c: middle right panel), and the ``dancing ghosts'' (b: bottom panels). The bottom panels include a view of PKS 2130-538 from the full-band image (leftmost) and the first sub-band image SPW:1, covering 36 MHz (middle) for a visual comparison of the sensitivity in both imaging settings. Also included in the bottom rightmost panel is a spectral index map of PKS 2130-538, generated with the first six sub-band images. 

\begin{table*}
    \centering
    \begin{tabular}{ccccccc}
    \hline
    \hline
         \textbf{SPW:2} & \textbf{SPW:3} & \textbf{SPW:4} & \textbf{SPW:5} & \textbf{SPW:6} & \textbf{SPW:7} & \textbf{SPW:8}   \\
         \hline
         99.6\% & 99.7\% & 99.7\% & 99.3\% & 99.5\% & 99.2\% & 98.9\% \\
        \hline
    \end{tabular}
    \caption{
    The percentage of the pixels in the error maps between $\sigma_u$-AIRI and $\sigma_s$-AIRI images with values below the estimated standard deviation of the noise in the image domain, $\sigma$, for each sub-band image of the field SB9351-12.
    }
    \label{tab:percentages}
\end{table*}

\begin{figure*}
\centering
\includegraphics[width=\textwidth]{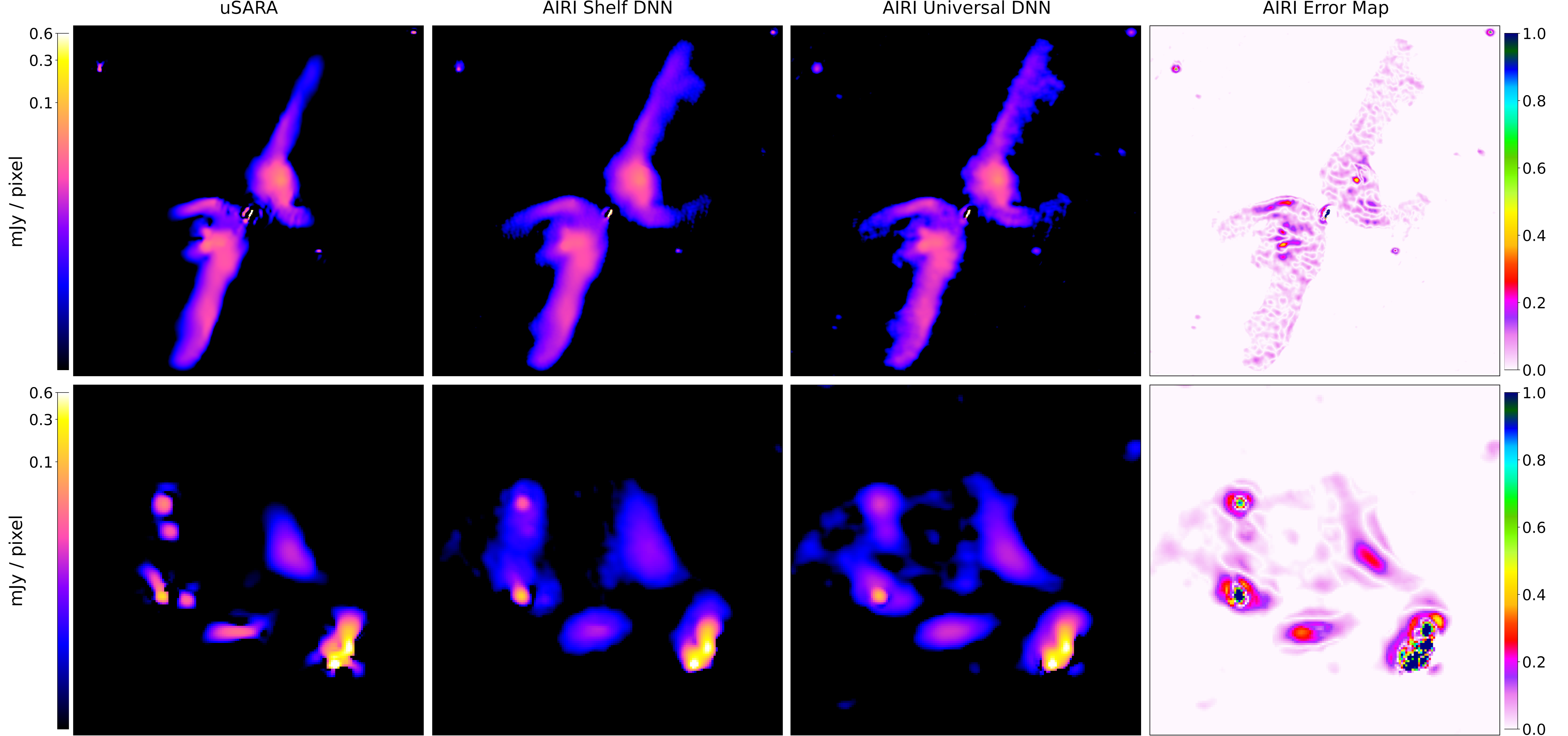}
\caption{Comparison of uSARA and AIRI reconstructions of the seventh sub-band data (SPW:7) of the field SB9351-12, focusing on the X-shaped radio galaxy (top) and the galaxy cluster SPT2023 (bottom). From left to right: uSARA reconstruction, $\sigma_s$-AIRI reconstruction (using the shelf-appropriate DNN denoiser),  $\sigma_u$-AIRI reconstruction (using a universal DNN denoiser), and the error map between $\sigma_s$-AIRI and $\sigma_u$-AIRI, normalised by the estimated standard deviation of the noise in the image domain, $\sigma$.} \label{variations}
\end{figure*}


\subsubsection{The Dancing Ghosts}
The separation between the curling secondary jet and the eastern lobe of the primary AGN in PKS 2130-538 is less distinct in our AIRI image when compared to the uSARA image from Part I. However, there is a more drastic difference in the improvement of resolution when moving to the full-band with AIRI. Our AIRI sub-band image shows a much smoother structure than the full-band image, particularly noticeable when focusing on the sharpness of the jet bridge linking the two lobes from the primary AGN. Faint point sources emerge more clearly in the AIRI full-band image, with a slight improvement over its uSARA counterpart. 

The filamentary emission extending from the eastern lobe (possibly a synchrotron thread similar to those discovered in \citealp{2020A&A...636L...1R}) appears slightly fainter and more diffuse in the AIRI image, with overall steeper spectra ($3.5 \leq \alpha \leq 6$) than seen in the uSARA maps. It is interesting that this eastern extending filament has such an ultra-steep spectrum in the AIRI map since the spectral index over the rest of the source morphology remains similar between the uSARA and AIRI maps. When comparing the flux density measurements in Table~\ref{tab:fluxghosts} to the corresponding measurements in Part I, there appears to be a clear consistency between uSARA and AIRI and {\tt WSlean}, with AIRI recovering slightly less flux at the higher frequencies. This source arguably has the most consistency in flux density measurements across the three different imaging methods, perhaps due to the overall flatter spectral index of the source and the lack of strong calibration artefacts in this field.

\subsection{Universal Denoiser and model uncertainty\label{ssec:results-strat}}

In an experiment toward modelling epistemic uncertainty, we measure differences between AIRI reconstructions of the field SB9351-12 produced by the two denoiser selection strategies proposed in Section~\ref{ssec:denoise-strat}, namely the denoiser shelf and universal denoiser strategies. The two approaches differ by the denoiser instance used. The AIRI reconstructions leveraging a pre-trained shelf of denoisers were presented in Section~\ref{ssec:SB9351}. Here we also present the reconstruction results when utilising the universal denoiser approach, and study the robustness of AIRI reconstructions to denoiser (\emph{i.e.}~model) variations. We recall that the considered universal denoiser corresponds to the lowest training noise level on the shelf, $\sigma_u=2\times 10^{-5}$. Under this consideration, we note that the first sub-band (SPW:1) is not included in this analysis since its shelf-appropriate denoiser corresponds to the universal denoiser (\emph{i.e.} $\sigma_s=\sigma_u$); therefore, only spectral windows 2--8 
were re-imaged. 

Focusing on our target sources of interest in this field -- namely, the X-shaped galaxy and the merging galaxy cluster SPT2023 -- reconstruction results of SPW:7, obtained using uSARA and the two AIRI denoiser selection strategies (the nearest shelf-appropriate DNN with $\sigma_s=8\times 10^{-5}$
and the universal DNN denoiser with $\sigma_u=2\times 10^{-5}$) are showcased in Figure~\ref{variations}. The seventh spectral window is chosen for a visual comparison due to its high signal and lower dynamic range when compared to other spectral windows. All other spectral windows imaged via the AIRI universal denoiser strategy are provided as FITS files and combined into a GIF for easier viewing in \citet{askapdataset}. In what follows, we refer to the AIRI reconstructions generated via their associated denoiser strategy as $\sigma_u$-AIRI (universal approach) or $\sigma_s$-AIRI (shelf approach). 

The most evident visual difference in the AIRI reconstructions -- particularly noticeable for the X-shaped galaxy -- is in the smoothness of the emission recovered in $\sigma_s$-AIRI and the arguably more detailed emission recovered in $\sigma_u$-AIRI which targets higher dynamic ranges. Both AIRI images recover significantly more diffuse emission than uSARA, yet with a slight compromise in resolution, as can be seen in the radio galaxies of the merging galaxy cluster SPT2023.

We examine the absolute difference between the $\sigma_u$-AIRI and $\sigma_s$-AIRI images. The resulting error map\footnote{A hard thresholding operation is applied to the error map, keeping only values above $10^{-8}$.} of SPW:7 is displayed in Figure~\ref{variations}, following a normalisation by the associated noise estimate in the image domain $\sigma$ (see Eq.~\ref{eq:heuristic}). From the full-FoV error maps associated with the sub-band images, we conduct a numerical analysis of AIRI reconstructions on the basis of the percentage of the pixels with values above $\sigma$. As shown in Table~\ref{tab:percentages}, for each sub-band error map, we find that $0.8 - 1.1\%$ of the pixels are of values higher than $\sigma$. This very small percentage corresponds mainly to pixel intensities within the brightest point-like sources. Because the integrated flux densities of these bright sources recovered by both AIRI reconstructions are very close, the discrepancy most likely arises from differences in individual pixel intensities which are slightly spatially offset from each other. 

When comparing the uSARA reconstruction to each of the AIRI reconstructions, we find that the percentage of the pixels with absolute difference above $\sigma$ is slightly more significant, at $2.5\%$ for SPW:7. These findings suggest that uSARA and AIRI reconstructions are very similar with respect to each other and that AIRI is highly robust to variations of the denoiser instance used (highlighting a small epistemic uncertainty of the learned denoiser approach). This also validates the simpler high dynamic-range universal denoiser strategy for AIRI, as opposed to the denoiser shelf approach.

\section{Computational performance}\label{sec:time}
For all AIRI imaging experiments performed in this work, the decomposition of the measurement operator and consequently the number of CPU cores allocated to enforce data fidelity are identical to uSARA (see Tables 7--9 in Part I for further details). While uSARA was deployed on the CPU nodes of Cirrus, AIRI was run on its GPU nodes comprising both CPU cores and GPUs  (see Section~\ref{sec:data} for details of the compute nodes). In this setting, the computing time of the forward step in AIRI was found to be up to 1.2 faster than its counterpart in uSARA, which we attribute to the newer processors used in these GPU nodes. 

The faceting functionality of AIRI was enabled, whereby the image is decomposed into $F=4$ facets. Hence, four GPUs were deployed for the parallel application of the DNN denoiser on each image facet. As such, the learned denoiser brought a drastic reduction of the computing time of AIRI's backward step by a factor 10 to 30 (depending on the image dimensions) compared to its pure optimisation counterpart uSARA. For each imaging experiment of each field, AIRI's total compute time and computational cost in CPU core hour are reported in Table~\ref{tab:time-airi}. In light of its extremely fast denoiser, AIRI's computational cost, solely driven by its forward step, is on average four times lower than uSARA (see Tables~7--9 in Part I) and five times higher than {\tt WSClean} (see Table~10 in Part I). 

Interestingly, preliminary experiments further leveraging GPUs to perform the Fourier Transforms involved in the forward step have shown a reduction of AIRI's total compute time, and consequently its computational cost by nearly a factor of two. However, a similar consideration in uSARA or {\tt WSClean} would not necessarily bring significant acceleration. The computational cost of uSARA would still be driven by its sub-iterative denoiser. Similarly, the Fourier transforms are typically not the dominating the computational cost of {\tt WSClean}. 

The speed and learning power of the DNN denoisers, pre-trained independently of the RI data under scrutiny, are significantly narrowing the gap between optimisation-based imaging algorithms and the standard CLEAN-based imager, thus highlighting their prospective potential for scalability and computational efficiency when handling extreme data and image dimensions.

\begin{table*}
   \begin{tabular}{cccccccccc}
\hline
\hline  
& & \textbf{SB8275}   &  & &\textbf{SB9351} &    & &\textbf{SB9442}    &\\
\hline
& $F$ & C\textsubscript{Image}& T\textsubscript{Image}&$F$ &C\textsubscript{Image}&T\textsubscript{Image}&$F$&C\textsubscript{Image} & T\textsubscript{Image} \\
                & & [CPUh] & [h] & & [CPUh] & [h] & & [CPUh] & [h] \\
    \hline
\textbf{Full-band} &  -- & -- & -- & -- &   -- & -- &  4 & 203 & 1.1 \\ 
\hline
\textbf{SPW:1}  & 4 & 48 & 1 & 4 & 95 & 2.1   & 4  & 45 &  1.4 \\ 
    \hline
\textbf{SPW:2}  & 4 & 63 & 1.1  & 4  & 95 & 2.2  & 4  & 54 &  1.5 \\ 
    \hline
\textbf{SPW:3}  & 4 & 104 & 1.4  & 4  & 101 & 2.6 & 4 & 51  & 1.6 \\
    \hline
\textbf{SPW:4}  & 4 & 104 & 1.6  & 4  & 105 & 2.7 & 4 & 50  &  1.6 \\ 
    \hline
\textbf{SPW:5}  & 4 & 119 & 1.7  &  4 & 99 & 2.3 & 4  & 52  &  1.5 \\ 
    \hline
\textbf{SPW:6}  & 4 & 129 & 1.7  &  4 & 104 & 2.5 & 4  & 62  &  1.6\\
    \hline
\textbf{SPW:7}  & 4 & 103 & 1.5  &  4 & 125 & 2.9 & 4  & 66 &  1.6\\ 
    \hline
\textbf{SPW:8}  & 4 & 144 & 2    &  4 & 91  & 2   & 4  & 69  &  1.6 \\ 

    \hline
    \end{tabular}
    \caption{AIRI computational costs for all imaging experiments: $F$ refers to the number of image facets, each deployed on one GPU core; C\textsubscript{Image} [CPUh] is the computational cost of the deconvolution in CPU core hours and T\textsubscript{Image} is the time in hours of the deconvolution. Memory and computational costs specific to the measurement operator are identical to those considered in uSARA (see Tables 7--9 in Part I for more details).
\label{tab:time-airi}}
\end{table*}


\section{Conclusions} \label{sec:con}

The results of this work show that the PnP-RI image reconstruction algorithm AIRI is on par with the precision-capability of its pure optimisation counterpart uSARA and surpasses the precision and robustness capabilities of {\tt WSClean}. A main and consistent feature of AIRI reconstructions is their sensitivity to the diffuse components of faint emission. This gives AIRI a distinct advantage over uSARA when the scientific goal is to detect and fully reconstruct low-surface-brightness diffuse emission at or near the noise level.

Building a shelf of suitable denoisers covering a range of potential target dynamic ranges has proven to be a solid approach for implementing AIRI. When resorting to a single universal high dynamic-range denoiser, high-fidelity reconstruction is also achieved, with comparable results to nearest-on-the-shelf reconstructions. In fact, we find that AIRI realisations reconstructed from denoisers with different training noise levels have about a 1\% discrepancy in terms of the percentage of pixel intensities above the estimated noise level of the imaged data.

In comparing the flux density values of the scrutinised sources reported in Parts I and II, we observe varying levels of consistency between uSARA and AIRI. A strong agreement is found in the case of the source `` dancing ghosts'', likely due to its relatively flat spectra and absence of strong calibration errors at its vicinity. However, when at least one of these conditions is not met, the flux density values of the other sources compare differently, yet with consistent variations across sub-bands. This is expected as AIRI has been shown to capture more faint and diffuse emission with slightly different morphology than uSARA. Concerning {\tt WSClean}, its flux measurements are generally higher than both AIRI and uSARA, particularly for the faint sources whose brightness is near or at the noise level, indicating possible over-estimation of the measurements taken from the restored images. Since each sub-and was imaged separately, one cannot assert that the measurements of one method are more reliable than the other, particularly since these early ASKAP observations have not been validated against standard flux catalogues. Wide-band variants of uSARA and AIRI are expected to provide more accurate measurements in the future.

For the analysed ASKAP FoVs, AIRI has demonstrated a four-fold acceleration on average over its pure optimisation counterpart, uSARA.  The higher computational efficiency achieved by AIRI is attributed to its substantially faster denoiser, and is a clear indication of its scalability power to large image dimensions. An in-depth investigation of practical scalability to extreme data dimension is warranted. The study will appear as part of on-going work towards a professional parallel C++ implementation of AIRI.

\section*{Acknowledgements}
The first two authors contributed equally to this work. This work was supported by the UK Research and Innovation under the EPSRC grants EP/T028270/1 and EP/T028351/1, and the STFC grant ST/W000970/1. The research used Cirrus, a UK National Tier-2 HPC Service at EPCC funded by the University of Edinburgh and EPSRC (EP/P020267/1). ASKAP, from which the data under scrutiny originate, is part of the Australia Telescope National Facility managed by CSIRO. This project used public archival data from the Dark Energy Survey (DES).

\section*{Data Availability}
The ASKAP data underlying this article (calibrated visibilities and mosaic images of Scheduling Blocks) are made publicly available for viewing and download on the \href{https://data.csiro.au/collections/#domain/casdaObservation/search/}{CSIRO ASKAP Science Data Archive} (CASDA; \citealp{2017ASPC..512...73C}), and can be accessed with the unique Project Identifiers AS034 and AS101. The reconstructed images in FITS format as well as the GIF files showing the imaged fields over the spectral windows are made available in \citet{askapdataset}. The uSARA and AIRI code will become available in a later release of the Puri-Psi library for RI imaging.


\bibliographystyle{mnras}
\bibliography{ASKAP-AIRI} 





\bsp	
\label{lastpage}
\end{document}